# COSMIC RAYS AND SOLAR INSOLATION AS THE MAIN CONTROL PARAMETERS OF THE CATASTROPHE THEORY OF CLIMATIC RESPONSE TO ORBITAL VARIATIONS

*V.D. Rusov[1], A.V. Glushkov[2], V.N. Vaschenko[3], V.N. Pavlovich[4], T.N. Zelentsova[1], O.T. Mihalus[1], V.A. Tarasov[1], D.N. Saranuk[1]*

[1]National Polytechnic University, Schevchenko av. 1, 65044 Odessa, Ukraine
[2]Odessa State Ecological University, Lvovskay str. 15, 65038, Odessa, Ukraine
[3]Ukrainian Antarctic Center, Peremogi av.10, 01033 Kiev, Ukraine
[4]Institute for Nuclear Research, of National Academy of Science, Nauki av. 47, 03028 Kiev, Ukraine

The energy-balance model of global climate, which is taking into account a nontrivial role of solar and galactic protons, is presented. The model is described by the equation of fold catastrophe relative to increment of temperature, where the variation of a solar insolation and cosmic rays are control parameters. It is shown that the bifurcation equation of the model describes one of two stable states of the climate system. The solution of this equation exhibits the property of the determined bistable behavior of climate at the global level and the possibility of appearance of the determined chaos of "the weathers" at the local levels.

The results of the comparative analysis of the computer simulation of the time-dependent solution of energy-balance model of global climate and the oxygen isotope records for deep-sea core V28-238 over the past 730 kyr are presented, and the evolution of climate on 100 kyr forward is also predicted.

It is shown that the proposed model successfully explains the nicety of the paleoclimatic records. The model is clear of all known difficulties of the Milankovich theory for the analysis and the interpretation of physical mechanism, by which the climate system responds to orbital forcing.

---

Corresponding author: Prof. Rusov V.D., E-mail: siiis@te.net.ua

According to the Milankovich astronomical theory of climate variations in the eccentricity of the Earth's orbit are the fundamental cause of the succession of Pleistocene ice ages [1-3].

However, the Milankovich theory not only successfully explains the nicety of the paleoclimatic records [4], but at the same time it is confronted by serious difficulties for the analysis and the interpretation of the physical mechanism, by which the climate system responds to orbital forcing. At first, the variations of solar heating are sufficiently small to control by the climatic cycles and practically do not exhibit intermediate (with period ~ 400 thousand years) and all the more long-time (with period ~ 2.5 million years) modulations, which are concerned with the periods of the eccentricity characteristic changes [5,6]. Secondly, the amplitude of one of the prominent interglacial events, isotope stage 11, has been large at times (around 400 kyr ago and today [7]), when the eccentricity modulation has been zero or, that is equivalently, the insolation variations are the smallest. This conflict is also called the " stage-11 problem " [7]. In third, improved measurements have uncovered an apparent causality problem: the sudden termination of glacial-interglacial cycles appear to precede the increases in insolation [8]. In addition, the Milankovich astronomical theory of climate, explaining Pleistocene ice ages, does not explain their absence in the majority of other geological epochs, though the insolation variations similar to the Quaternary, apparently, had a place always [9].

The use of so called "flip-flop back and forth" mechanism to set up the models of climatic response to orbital variation [8] was the first strategic hitch on the path to overcoming of these difficulties. The brightest implementation of such a strategy was the original work on stochastic resonance by Benzi et al., in which the term was coined [10]. The principal manifestation of the phenomenon is strong different output characteristics of the system to a weak periodic signal; this reaction grows with increasing of the noise level up to certain extent. The main results point to the possibility of explaining large amplitude, long-term alternations of temperature by means of a co-operation between small external periodic forcing due to orbital variations (about 0.1 % of the solar constant) and an internal stochastic mechanism (for example, a white noise, which simulates the global effect of relatively short-term fluctuations in the atmospheric and oceanic circulations on the long-term temperature behavior [10]). Note that similar results have been obtained by Nicolis [11] using a different approach.

The original work on deterministic transitions in multi-state climate model by Paillard [13] is the alternate branching of "flip-flop"-strategy application. In this work it is supposed that depending on the insolation forcing and the ice volume, the climate system can enter three different regimes called **i** (interglacial), **g** (mild glacial) and **G** (full glacial), between which phase transitions in **i-g-G-i** direction take place at given boundary conditions. The three-state model is not vary sensitive to the initial conditions and can be a good candidate to explain some puzzling features of the Quaternary records, in particular the "100 kyr problem" as well as "stage-11 problem" [13]. This work grows out of the generalization of successful ideas and achievements, which were obtained during the difficult, but effective development of "ice-volume" models of climatic response to orbital variation by Calder [14], Weertman [15], Paillard [16] and Imbrie J.-Imbrie J.Z. [8].

But it is necessary to note, that despite the reached progress in the description of multi-state climate, the general disadvantage of the works of "flip-flop"-direction is the fact that the stable states of the earth's climatic system [10,13] and transitions between them [13], figuratively speaking, "are set by hands".

In this sense Sellers-Byduko-type models (so-called heat-budget direction [17-19]), in which was made a preliminary attempt verifying the Milankovich theory using a zonally symmetric energy-balance climate model forced with seasonally varying insolation (as function of latitude and time of year) and obtain generally favourable results [20, 21], are no exception.

We consider that all problems of the modern energy-balance climate models caused by the necessity to use strong assumptions are the direct consequence of the traditional neglect of the role of the cosmophysical factors in the formation of global climate of the Earth (in particular, the influence of solar and galactic cosmic rays (GCR) on climate).

The opponents of the account of the solar activity influence on meteoparameters raise the objections, which have energy nature. The atmospheric processes are characterized by power about $10^{26}$-$10^{27}$ erg/day. At the same time the energy entry from the solar wind to the magnetosphere and the subsequent processes in the magnetosphere and the ionosphere is about $10^{23}$ erg/day. It is less than the power of real atmospheric processes on 3-4 orders. Therefore, the typical and settled conclusion follows: " … the relative contribution from cosmic rays to the flow of solar radiation is very small and they do not directly cause any essential variations of the weather and climate. And so, there is no wonder that variations with the period of solar cycle about 11.5 years are not observed in the spectra of climatic series" [9].

However, as it was shown by Pudovkin et al. [22,23], another explanation of considered problem is possible. In spite of the fact that the energy contribution from solar and galactic cosmic rays is not comparable with the value of solar insolation, they effectively influence on physico-chemical processes in the atmosphere by the stimulation of nitrogen and hydrogenous cycles of physico-chemical reactions and considerably change the optical properties (transparency) of the atmosphere. It means, that the variations of cosmic rays play a part of the peculiar physical modulator of additional solar energy incoming to the lower atmosphere. Long correlation experiments and theoretical estimations [22,23] have shown that the additional (solar) energy incoming to the atmosphere at change of its transparency due to the perturbation caused by cosmic rays has approximately same value as the energy of processes exited in the atmosphere. The independent data of Raisbeck et al. [24], which indicate a high correlation score in time between the concentration of cosmic beryllium ($^{10}$Be) accumulated in ice and the concentration of heavy isotope of oxygen ($\delta^{18}$O) in ice core obtained at the station "Vostoc" (the specific analog of paleotemperature), can serve as an illustration to above said (Fig. 1). Thus, the account of the additional energy caused by the change of the atmosphere transparency results in the following energy-balance equation of the Earth climate system (ECS) [25]:

$$\Delta U(T,t) = P(t) \cdot [1-\alpha(T)] - I(T) + G(T,t), \qquad (1)$$

where the first member of Eq. (1) $\Delta U(T,t)$ characterizes the additional power of heat generation in ECS relative to initial power $U_0 = U - \Delta U$, W; $T$ is temperature, K; $t$ is the time, for which an energy balance is considered; $P(t)$ is the flux of solar radiation on the upper bound of the atmosphere, W; $\alpha$ is ECS albedo; $I(T)$ is the intensity of long-wave heat radiation of the atmosphere outgoing from the upper bound of the atmosphere to cosmos, W; $G(T, t)$ is the additional flux of solar heat in ECS generated by the flux of galactic and solar cosmic particles, W.

Let us express energy components *I*, α and *G* of this equation as the function of temperature. The first energy term *I* corresponding to long-wave radiation with the average temperature *T* in sufficient for our purpose approximation is equal :

$$I = \gamma \sigma T^4, \qquad (2)$$

where σ is Stefan-Boltzmann constant, γ is the area of the external boundary of the upper atmosphere.

The temperature dependence of the effective value of ECS albedo, which, in essence, determines the quantity of reflected direct solar radiation in ECS, is selected as Faegre continuous parametrization [26]:

$$\alpha = 0.486 - \eta_\alpha \cdot (T - 273), \quad where \quad \eta_\alpha = 0.0092 K^{-1}. \qquad (3)$$

At last, let us consider the problem of the functional dependence of additional heat flux *G* on temperature. It is known that the integral GCR spectrum *N(E)* is surprising stable and in the range of $10^{11}$-$10^{15}$ eV is power-mode:

$$N(E) \sim E^{-\mu}, \quad \mu \approx 1.7. \qquad (4)$$

Let us suppose that the GCR energy is completely absorbed in the atmosphere. Then we can estimate the average energy $E_g$ transmitted to atmospheric gas:

$$E_g \sim NE \sim E^{1-\mu}. \qquad (5)$$

It is correctly to consider that each cosmic particle induces in moving gas medium the production of a gas vortex with size λ, which is inversely proportional to particle energy *E*, i.e., $E \sim \lambda^{-1}$.

Following Ref. [27], let us replace the scales λ by the corresponding "wave numbers" of vortex pulsations in the form of $k \sim 1/\lambda$. Then taking into account Eq. (5) the integral spectrum $E_g$ of gas vortexes looks like

$$E_g \sim k^{1-\mu}, \qquad (6)$$

that corresponds to energy spectral density

$$E_g(k) \sim k^{-\mu}, \qquad (7)$$

where $E_g(k)dk$ is the kinetic energy of gas vortex with spatial wave number $k$.

Since $\mu \approx 5/3$, it is obvious that the spectrum (7) is nothing else than known Kolmogorov-Obukhov spectrum [27,28], which describes the dynamics of high-frequency perturbations or, in other words, the structure of small-scale turbulized medium in the form of the vortex cluster skeleton with fractal dimension $D=5/3$ [27,28].

Note that the appropriate laws of similarity, scale correlations and spectral dynamics (in particular, within the framework of the inertial interval theory reducing to the Kolmogorov-Obukhov vortex energy spectrum) are traditionally applied for the simulation of atmospheric turbulence in boundary layer, i.e., for the layer of the air, within the limits of which the interaction of the atmosphere and an underlying surface is directly exhibited [28].

Hence the obtained Kolmogorov-Obukhov spectrum (7) induced by GCR has differences not only in the reason, but also in the main place of its formation: the homosphere-the upper atmosphere. It is necessary to mention here known experimental data [29], which are confirm the existence of small-scale turbulence in this area of the atmosphere. On the other hand, if the large-scale fractal structure (in the form of the skeleton of vortex cluster) is formed in this area, the energy can be output to "infinity" through it, i.e., the energy can be output from the area of turbulent motion, for example, to the upper atmosphere. Then the question about what experimentally observed consequences may be in this case arises.

The universal regularities obtained within the frameworks of the inertial interval theory (or, in other words, Kolmogorov-Obukhov laws of similarity [27,28]) were developed by Obukhov for the description of the statistical structure of the temperature turbulent pulsations, when they do not yet influence essentially on flux structure [30]. Moreover in Ref. [30] it was shown that the structure of the temperature field in turbulent conditions is determined not only by the dissipation rate of turbulence kinetic energy per mass unit $\varepsilon$, but also by the dissipation rate of temperature fluctuation intensity $N_T$, which is equal on order of values:

$$N_T \cong (\Delta T)^2 \cdot \Delta u \cdot L^{-1}, \qquad (8)$$

where $\Delta u$ is characteristic size of the scale of the rate of main energy-carrying vortexes; $\Delta T$ is characteristic temperature difference in a flux on its external scale $L$.

Then it is possible to present the generalized Kolmogorov-Obukhov law, which take into account both the turbulent pulsation of kinetic energy and temperature pulsation, in equivalent spectral (on space) form. Let $E_T(k)dk$ is the kinetic energy of unit liquid mass contained in pulsations with values $k$ in given interval $dk$. As the function $E_T(k)$ has dimension of cm$^3$/s$^2$, making the combination of this dimensionality from $N_T$, $\varepsilon$ and $k$, we obtain in the inertial interval of scales the following expression:

$$E_T(k) = C_{1T} N_T \varepsilon^{-1/3} k^{-5/3}, \quad C_{1T} \approx 1.4, \qquad (9)$$

Integrating expression (9) in an interval $k \in [k_L, \infty]$ and taking in account the order of values of energy dissipation $\varepsilon \sim (\Delta u)^3/L$ and the intensity of temperature fluctuations (8), we have

$$E_T = C_{1T} \frac{N_T}{\varepsilon} \cdot \left(\frac{\varepsilon}{k_L}\right)^{2/3} = C_{1T}(\Delta T)^2. \qquad (10)$$

Analyzing the expression (10) it is possible to write down without loss of quality the general form of the dependence of heat flow $G_T$ in turbulent conditions on characteristic temperature difference $(T_{out}-T)$ in the flux on its external scale $L$:

$$G_T \sim g \cdot (T_{out} - T)^2. \qquad (11)$$

Let us remind that in our conception the variations of additional insolation described by parameter $g$ (W/K$^2$) are determined mainly by the intensity of solar proton events in the atmosphere. On the other hand, it is obvious that the intensity of proton events in the atmosphere is modulated by an eccentricity $e(t)$. Thus, it is possible to suppose that the generation rate of additional insolation transported to the upper atmosphere by the turbulent mechanism of heat transfer in linear approximation is proportional to the intensity of solar proton events in the atmosphere or, in other words, to the value of an eccentricity $e(t)$:

$$g(t) = k \cdot e(t), \qquad (12)$$

where $k$ is constant coefficient with dimensionality of W/K$^2$, which in the general case should depend on latitude in zone ECS.

At last, substituting all partial contributions of heat flows (2), (3) and (11) to resulting energy-balance expression (1), we obtain

$$-\frac{1}{4\gamma\sigma}\Delta U(T,t) = \frac{1}{4} \cdot T^4 + \frac{1}{2} a(t) \cdot T^2 + b(t) \cdot T + absolute\ term, \qquad (13)$$

$$a(t) = -g/4\gamma\sigma, \quad b(t) = -\eta_\alpha P/2\gamma\sigma. \qquad (14)$$

It is obvious that the expression (13) describes the collection of energy-balance functions $\Delta U(T,a,b)$, which depend on two control parameters $a(t)$ and $b(t)$. There is no difficulty to note, that this collection represents so-called "potential" of fold catastrophe [31,32] (Fig. 2). In future we will be interested by an equation of fold catastrophe (13) relative to increment $\Delta T = T - T_0$ of the following type:

$$\Delta U(T_0 + \Delta T, a, b) - \Delta U(T_0, a, b). \qquad (15)$$

where $T_0$ is the average temperature averaged on the respective time interval $\Delta t$.

Omitting the nicety of the computational methods of catastrophe theory [31,32] we show only that, using the remarkable lemma of Morse [31,32], with the help of diffeomorphism (smooth local replacement of coordinates) it is possible to obtain the increment for the first term in the second member of Eq. (13) in the following equivalent form

$$(T_0 + \Delta T)^4 - T_0^4 \cong 5 \cdot 10^{-3} \cdot T_0^3 \cdot (\Delta T)^4 + 4 \cdot T_0^3 \cdot \Delta T, \quad for \quad \Delta T = 0 \div 4K. \qquad (16)$$

Let us remind that the normalized variation of insolation $\Delta \hat{S}$ with average $<\Delta \hat{S}>=0$ and dispersion var($\Delta \hat{S}$)=$\sigma^2_{\Delta \hat{S}}$=1 is applied more often for the simulation of the ECS :

$$\Delta \hat{S} = \frac{\Delta S - \langle \Delta S \rangle}{\sqrt{\text{var}(\Delta S)}}, \quad where \quad \Delta S = S - S_0, \qquad (17)$$

where $S_0=P(t=0)/\gamma$ is specific solar insolation (W/m²) in the present point of time $t=0$, $S=P(t)/\gamma$ is specific insolation (W/m²) in the point of time $t$, $\gamma$ is the area of the external boundary of the upper atmosphere.

Transforming Eq.(13) according to computing circuit (15) and taking into account Eqs.(16)-(17), we obtain the following expression for the increment of additional power $\Delta U$ relative to temperature disturbance $\Delta T$:

$$V(\Delta T, t) = \frac{1}{4}\Delta T^4 + \frac{1}{2}a(t)\cdot \Delta T^2 + b(t)\cdot \Delta T + absolute\ term, \quad V = -\frac{10^2}{2\sigma T_0^3}\Delta U, \qquad (18)$$

$$a(t) = -a_0 \cdot e(t), \qquad (19)$$

$$b(t) = -b_0 \cdot \left[\Delta \hat{S}(t) + \xi_G\right], \qquad (20)$$

where

$$a_0 = \frac{100k}{\gamma \sigma T_0^3}\left[K^2\right], \quad b_0 = \frac{\eta_\alpha \sigma_{\Delta S}}{2 \cdot 10^{-2}\sigma T_0^3}\left[K^3\right], \quad \xi_G = \frac{1}{\sigma_{\Delta S}}\left[\langle S \rangle - \frac{4\sigma T_0^3}{\eta_\alpha}\right] \geq 0. \qquad (21)$$

The physical sense of the inequality $\xi_G \geq 0$ consists in the fact, that its unsatisfaction results in the absurd result: the growth of the average temperature $T_0$ results in a cool-down (as it follows from Eq. (20)).

Sufficiently nonsimple behavior of the "potential" function $V(\Delta T, t)$ can be to represented by the united geometrical picture (Fig. 2), which permits to show the variety of catastrophe or, in other words, the surface of equilibrium in three-dimensional space with coordinate axes $\Delta T$-$a$-$b$ [31]. The canonical form of the variety of the fold catastrophe, which represents point set $\Delta T$-$a$-$b$, satisfies the equation:

$$\nabla V(\Delta T, t) = \Delta T^3 + a(t)\cdot \Delta T + b(t) = 0. \qquad (22)$$

The Eq. (22) looks like a surface with fold (Fig. 2a). The projected fold as the line of a tuck (the set of critical points) is shown in Fig. 2b. Note that to the neighborhood of each point of the surface there locally corresponds its curve of "potential" (18) on ($a,b$). Some of such dependences are shown in the mapping of the catastrophe (Fig. 2c).

The bifurcation set of fold catastrophe (or the set of critical points $a_c(t)$ and $b_c(t)$ depends on the nature of the roots of Eq. (22), i.e., on discriminant $D=4a^3+27b^2$, and therefore it is described by the equation of semicubical parabola

$$[a_c(t)/3]^3 + [b_c(t)/2]^2 = 0. \qquad (23)$$

Thus the general bifurcation problem of the solution $\Delta T(t)$ determination is reduced to the determination of the solution set of Eq. (22) for the appropriate joint trajectory $a(t)$, $b(t)$ in the space of control parameters.

Now we are ready for the determination of the solution $\Delta T(t)$ of bifurcation problem (22), for example, for latitude 65° N. It is obvious, that base (18) and the bifurcation (22) equations within the framework of any zonal energy-balance climate model formally save their form, but only concerning the variations of control parameters $a(t)$, $b(t)$ at latitude 65° N. In its turn, control parameters $a(t)$, $b(t)$ "are control" by the variations of the main and additional insolation. The nicety of zonal climatic model are exhibited with the advent of an advection term in Eqs. (1) and (18), which characterizes the flows of explicit and latent heat through the lateral faces of the element (zone) of ECS in the form of complementary constant $A$ in expression $\xi_G$ for control parameter $b(t)$

$$\xi_G(A_{65°N}) = \frac{\langle S \rangle}{\sigma_{\Delta S}} - \frac{4\sigma T_0^3}{\eta_\alpha \sigma_{\Delta S}} + A_{65°N} = const. \qquad (24)$$

For the solution of bifurcation Eq. (22), which describes the extreme values of the increment of temperature $\Delta T$ for the element (zone) of ECS in latitude 65° N, it is necessary to determine the values of two climatic constants $k$ and $\xi_G(A_{65°N})$ in Eqs. (19)-(21) and (24). The values of remaining parameters (except for average temperature $T_0$) are known. For example, the values of an eccentricity $e(t)$ (Fig. 3a) and normalized insolation $\Delta \hat{S}(t)$ (Fig. 3b) are calculated by Berger [4-5]. The value of mean-root-square error of the variation of insolation is equal to $\sigma_{\Delta S}$=18.26 for the selected time period including 730 kyr in past and 100 kyr in future.

The selection of value of the average temperature $T_0$ was made from following considerations. The value of modern climatic representative temperature under the data Ghil [17,11] and Fraerdrich [11] is about $T$=288.6 K. On the other hand, according to the data of Russian-French-Italian researches of an ice core from a hole at Russian Antarctic station "Vostoc" [33] the average value of the range of increment $\Delta T$ = [2, -6] is less approximately on 2 K than the modern temperature. Therefore the value of average temperature equal to $T_0$=286.6K was used for the further calculations.

The traditional method of calibration relative to experimental data was applied to the determination of climatic constants $k$ and $\xi_G(A_{65°N})$. The essence of this method lies as follows. According to the experiments, which were made at Antarctic station "Vostoc" [33], it is possible to conclude that the jump of the temperature $\Delta T$ relative to average temperature $T_0$=286.6 K was approximately $\Delta T \cong 4K$ at $t$=120 kyr ago. Such a supposition with regard for Eq. (23) reduces to the following single-valued form of bifurcation equation (22):

$$\Delta T^3 - 12 \cdot \Delta T - 16 = 0, \quad for \quad t = 120 kyr. \tag{25}$$

Hence the fixed values of control parameters $a_{120}=12K^2$ and $b_{120}=16K^3$ at $t=120$ kyr in the past in latitude 65° N make it possible to determine the values of climatic constants $k$ and $\xi_G(A_{65°N})$ from Eqs. (19)-(21) and (24). Taking into account that at $t=120$ kyr in the past to the value of eccentricity $e_{120}=0.038$ there corresponds the value of the normalized variation of insolation $\Delta\hat{S}_{120}=2.3$, and also using the values of the standard error of the variations of insolation $\sigma_{\Delta S}=18.26$, Faegre parameter (3) $\eta_\alpha=0.0092$ K$^{-1}$ and Stefan-Boltzmann constant $\sigma=5.67 \cdot 10^{-8}$ W/m$^2$K$^{-4}$ we obtain the following values of two climatic constants $k$ and $\xi_G(A_{65°N})$ for the average temperature $T_0=286.6$ K:

$$(k/\gamma) = 4.20, \quad \xi_G(A_{65°N}) = 0.25. \tag{26}$$

The time-dependent solution of bifurcation equation (22), which for the given method of the calibration of the value of climatic constants $k$ and $\xi_G(A_{65°N})$ describes the temporal changes of the increment of temperature $\Delta T$ relative to the average temperature $T_0=286.6$ K for 730 kyr in the past and 100 kyr in future, is shown in Fig. 3d. In Fig. 3c oxygen isotope curve for deep-sea core V28-238 from Pacific Ocean over the past 730 kyr is presented. Data from Shackleton and Opdyke [34] are plotted against the PDB standard on the time scale of Kominz et al. [35]. The high goodness of fit between experimental (Fig. 3c) and theoretical (Fig. 3d) data is the peculiar indicator of the high quality of the prognosis of the temporal changes of global temperature $T_0+\Delta T$ in latitude 65°N over the next 100 kyr (Fig. 3d).

It is known, that the less number of parameters the better the quality of model other things being equal. Therefore it is necessary now to tell about the number of the model parameters. It is obvious, that the number of parameters in multizonal model is equal to the order of the set of equations (18), determined by the number of latitude zones, into which the hemisphere of the Earth is divided. On the other hand it is important to note that one-zone or, in other words, global climate model is practically parameterless, since the climatic constant $k$, $\xi_G(A=0)$ and the average temperature $T_0$ are completely determined by expressions (21) (taking into account the indicated above calibration relative to experimental data). In this sense, the expression "practically parameterless model " means only that in the global climate model the values of control parameters $a(t)$ and $b(t)$ are set not by "hands", but are obtained as a result of independent calculations.

At this point it is appropriate to put the equation if the cosmic rays and, first of all, solar protons, which play the part of the peculiar catalyst of the turbulent mechanism of additional solar heat transport to the upper atmosphere, can be the reason of essential variations of the paleointensity of geomagnetic field. The answer is affirmative and that is why.

The physical mechanism of change of the atmosphere optical transparency under the influence of cosmic rays predetermines a strong correlation between the maxima of solar proton intensities (predetermined by the maxima of eccentricity) and the largest changes of atmosphere albedo. Due to this strong correlation the solar and space protons should be not only the reason of perceptible variations of the paleointensity of geomagnetic field (Fig. 3e), but the periods of these variations should be close to the typical periods of the eccentricity amplitude. Strikingly, but it turned out, that this phenomenon really takes place and is steadily observed as equally assigned frequencies, corresponding to the typical periods (~100 and ~410 kyr) of the modulations of eccentricity amplitude, both in the spectrum of variations of the increment of temperature (Fig. 4b) and in the spectrum of variations of virtual axial dipole moment of the Earth (VADM) for time periods 0-800 kyr in the past [36] (Fig. 4a). Thus, it is possible to assert that the variations of the eccentricity influence independently both on magnetic field and climate of the Earth. However, it is necessary to take into account that the relative changes of true paleointensity will be different not only due to the different intensity of solar protons, but also due to the influence of varied conditions of sludging on a paleomagnetic record. It is obvious, that basing on the time-dependent change of the variations of global or zone temperature, a real opportunity to estimate quantitative influence of the climate changes determining the conditions of sedimentation on the record of paleointensity variations of terrestrial magnetic field appears.

Note that there are grounded notions about possible periodic variability of the Sun activity under the influence of convection caused by the variable gravitational field of the planetary system [37], and it should be reflected in the dynamics of paleoinsolation. The most interesting period in this model is the timing loop ~26 mill. years. For the first time cycle of such duration in geologic data was found out by Newall [38], who analyzed the data about the first and the last occurrence of 2250 families of animals of large groups in the paleontologic chronicle. Recently Raupe and Sepkoski (1988) carried out the similar research at more extensive sample of 9773 genera of zoolites for the period from Permian formation up to the end of the Pliocene and confirmed the presence of the strongly pronounced interval of the kind extinction, which was approximately 26 mill. years [38]. Such a regular periodicity is evidence of the fact that the extinction was caused by any astronomical factor, which is connected, for example, with the Sun, the Solar system and/or the Galaxy [38]. In addition, during these researches close correlation between the paleoclimatic characteristics and the geologic data relating to the biological evolution of the Earth was found [39,37]. This fact has increased the interest to this problem. For example, it was shown that the climate cooling and it humidisation lead to increase of the number of the organism families, whereas the warming and aridisation cause their decreasing [37].

There is no wonder that our mechanism of cosmic rays action on the variation of geomagnetic field and climate and the connection of these variations with the value of solar proton intensity and eccentricity causes the natural question: "Do some singularities of galactic protons manifest or their role is limited only by the role of active "witness?" The essence of the question becomes clear, if we will remind that the main source of space protons are supernova outbursts near to Galaxy center and therefore the intensity of galactic protons is directly determined by the corresponding elements of galactic orbit of the Solar system. It is interesting that in the resonance model of the galactic orbit of the Sun and its system [37] the following time intervals of orbital elements are noticeable: the interval of 50 million years corresponds to the period of the orbit variations relative to the plane of the Galaxy; the interval of 75 million years corresponds to the cycle of motions of the Sun between two points - apogalaxion and perigalaxion; the interval of 150 million years corresponds to the cycle of motion between apogalaxions (or perigalaxions) and interval of 225 million. years, i.e., the sidereal period of the Sun around of the Galaxy. It is obvious, that our model (18) through control parameter $a(t)$, such as Eq. (19) will feel changes of the galactic proton intensity caused, for example, by the variations of the orbit of Solar system with the period of ~50 million years relative to the plane of the Galaxy. These variations of the intensity of galactic protons cause the changes of magnetic field and climate of the Earth with the period equal approximately 25 million. years, which, in its turn, is explained by the cyclic passing of the orbit of Solar system through the plane of the Galaxy. Taking into account that the interval between the maxima on extinction curve of Raupe and Sepkoski [38] besides the period of 25 million years contains periods between the higher maxima, which are equal to 75 and 150 million. years, it is possible to suppose that such a coincidence between the periods of the variation of the orbital elements of Solar system (and, hence, of global climate of the Earth (18)) and the periods of change in the number of kinds of organisms on the Earth is hardly random.

On the other hand, the lifting of continents dated for these moments can be connected with the nature of galactic gravitational field action on the rotating Earth [37]. And the motion of continents (as Monin correctly notes [9]) causes, respectively, the change of the part of their area in an equatorial zone and polar regions. This, apparently, became the main reason of gradual decrease of the level of temperature background reliably established by the facts [9] or, in other words, of gradual decrease of the average temperature $T_0$ during the Cainozoe, i.e., at last 67 million years.

Such decrease of temperature and simultaneous appearance of glacial epochs in the Cenozoic many researchers connect with certain resonance properties of ECS, which, for example, is unresonance at the averaged temperature $13°$ C dominating up to a Cenozoic, and responds by resonance oscillations generating interglacial and glacial epochs at the averaged temperature below $10°$ C [9]. Such a hypothesis for the first time was advanced by P.Woldstedt [40] in 1954.

In this connection, let us consider the similar role of the absolute value of the average temperature $T_0$ within the framework of our model. With that end in view we shall launch into digression? which is necessary for the explanation of a special role of temperature background level in dynamics of ECS. It is obvious, that time-independent ($t=t_0$) probability density function $p$ of random increment $\Delta T$ for the potential of fold catastrophe (18) looks like

$$p(\Delta T, a(t_0), b(t_0)) = \frac{1}{Z_0} \exp\left(-\frac{V(\Delta T, a(t_0), b(t_0))}{D/\Delta t}\right), \qquad (27)$$

where $1/Z_0$ is normalization factor, $D$ is diffusion constant (the constant of "chaotic state" [31]), $\Delta t$ is the interval of averaging of temperature in Eq. (1).

It is easy to see that Eq. (27) represents the stationary solution of Focker-Plank equation:

$$\frac{\partial p}{\partial t} = \frac{\partial}{\partial(\Delta T)}\left(\frac{\partial V}{\partial(\Delta T)} p\right) - \frac{\partial}{\partial(\Delta T)}\left(D\frac{\partial p}{\partial(\Delta T)}\right), \qquad (28)$$

which describes the changes of the ensemble of climate states with initial distribution such as Eq. (27). This distribution arises, unlike the similar equation in Ref. [41], under the influence of not internal, but external random factors, i.e., control parameters $a(t)$ and $a(t)$.

In this connection let us note three moments, which are very important for the understanding of model possibilities. At first, the presence in Focker-Plank equation (28) right member of two terms, which are in charge of drift and diffusion, is evidence of the fact that it is an equation with two temporal scales. The rapid time scale $\tau_1$ connected with an inverse relaxation to local minimum after perturbation is predetermined by "drift" term, and the diffusion predetermines slow time scale $\tau_2$, connected with transition from the metastable minimum to global one.

Secondly, Gilmore [31] (and Kramers even earlier [42]) obtained the estimation of these temporal scales of Focker-Plank equation such as Eq. (28). For comparison let us show the formulas for time scale $\tau_1$ (relaxation to the local minimum) and $\tau_2$ (diffusion from a metastable minimum in global one):

$$\tau_1 = \frac{1}{\lambda_1}, \quad \tau_2 = \frac{2\pi}{|\lambda_1 \lambda_2|^{1/2}} \exp(\Delta V/D), \tag{29}$$

where $\lambda_1$, $\lambda$ is the curvature ($d^2V/dx^2$) of function in the local minimum and local maximum, respectively, $\Delta V = V(\Delta T_M) - V(0)$, $\Delta T_M$ and $\Delta T_m = 0$ are the coordinates of the metastable minimum and maximum of the potential of fold catastrophe.

In third, Gilmore has shown that the velocity of motion of critical points $\Delta T_c(t)$ of the potential function $V(\Delta T, c(t))$ is comparable to derivative $dc/dt$, and it made possible to formulate the applicability requirements of the known agreements of Thom and Maxwell [31,32] (which was intuitive up to that moment) using the time derivative of control parameters $c \in \{a(t), b(t)\}$:

$$\text{Maxwell's principle: } \tau_2^{-1} \gg dc/dt, \tag{30}$$

$$\text{Thom's principle: } \tau_1^{-1} \gg dc/dt \gg \tau_2^{-1}. \tag{31}$$

In order to obtain the periodic temperature variations in our model (18), the use of Thom's principle of maximum delay (31) is necessary as, unlike Maxwell's principle, it ensures the presence of hysteresis loop on bifurcation set in the plane of control parameters (Fig. 2). In Fig. 5 the pictorial representations of the local dependence of function $p(\Delta T)$ (which is generally bimodal) on control parameters ($a, b$) at the neighborhood of some characteristic points of the surface of variety of catastrophe (22) (see Fig. 2). It is obvious, that the increase of the average temperature $T_0$ automatically reduces to the increase of generalized parameter $\xi_G$ in Eq. (21). This, in its turn, meets following geometrical illustration: path of $c \in \{a(t)<0, b(t)<0\}$ in the space of control parameters mostly is in the second quadrant (Fig. 5), i.e., high temperature $T_0$ retains ECS in "hot climate" state ($a(t)<0, b(t)<0$) and ECS states corresponding to glacial periods ("cold climate", i.e., $a(t)<0, b(t)>0$). are completely excluded. This fact in natural way explains the occurrence of glacial periods in the Pleistocene and their absence in the majority of other geologic epochs, although the insolation variations similar to the Quaternary, apparently, took place always.

The analysis of bifurcation equation (22) shows that the modern thermic mode (which, in essence, is the modern version of so-called 11 stage of paleoclimate [7]) is very sensitive to hypothetical changes of solar constant. According to our calculations, decreasing of solar constant on 1.5 % may convert the existing thermic mode to "white Earth" condition. Such a conclusion at the qualitative level is confirmed by decreasing of generalized parameter $\xi_G$ in Eq. (21), whose value is modulated by the value of the average solar insolation $<S>$.

At last, it is necessary to say that the Earth climate system within the framework of given model (18) practically is not sensitive to the initial data. It means that the system has no restrictions on the global time horizon, i.e., its global behavior is predictable, and it contrary to the known hypothesis that climate should exhibit the properties of deterministic dynamics, which haves a chaotic attractor of small dimension [43]. As to the local weather forecasts, as it is known, the best of used now models have about $6 \cdot 10^6$ variables, and therefore the error of prediction doubles each three days. Nevertheless, the casual possibility of "peaceful" coexistence of the practically infinite global time horizon (climate) and the finite local time horizon (weather) can be illustrated by the example of the fractal portrait of the potential of fold catastrophe (Fig. 6), whose hierarchical structure includes the determined bistable behavior of climate at the global level and the possibility of the manifestation of the determined chaos of "weathers" at the local level. Note, the fractal nature of the phenomenon can be exhibited in the multizone version of the energy-balance model (18) of Langevin type, in which the local fluctuations (on short time intervals) play the prime role. In other words, the Langevin type of multizone energy-balance model (18) (in the form of the system of stochastic differential equations) under certain conditions can ensure such a type of non-linear dynamics of ECS, in which, according to figurative expression of M.Berry, "determinism, similarly to the English queen, reigns, but does not rule" [44].

In summary, it is possible to say, that the proposed energy-balance model of global climate, which takes into account the nontrivial role of solar and galactic protons, not only successfully explains many nicety of the paleoclimatic records, but also in a natural way overcomes all known difficulties of the Milankovich theory arising during the analysis and interpretation of physical mechanism, by which the climate system responds to orbital forcing. However, in our opinion, the most important result of the offered model is the statement that the global climate of the Earth, on the one hand, is completely determined by two control parameters - cosmic rays and solar insolation, and, on the other hand, it has practically no restrictions on global time horizon, i.e., it is quite predictable. On the grounds of this statement the possibility for the exact formulation of the nontrivial problem of creation of the theory and methods of the global climate management opens for the first time, at least on the level of fixation of the modern state of climate [31].

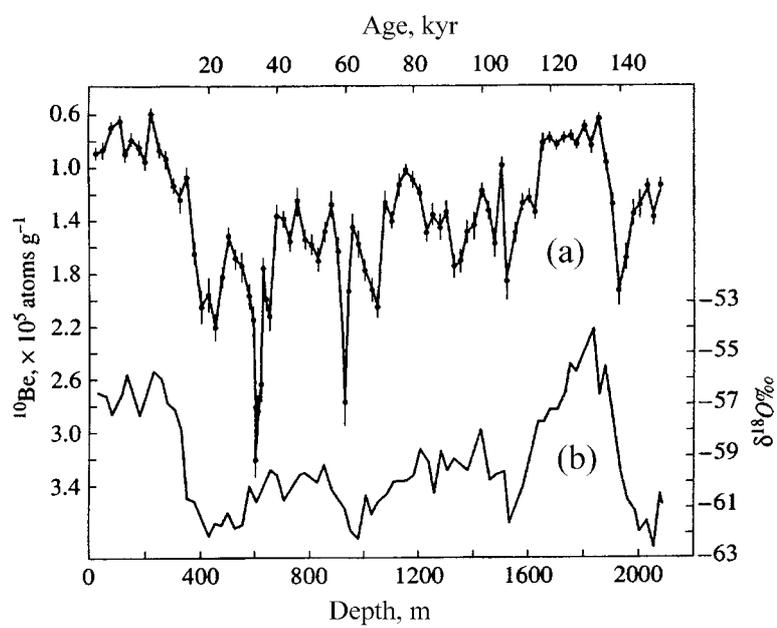

Fig. 1. The concentration of beryllium $^{10}$Be (a) and heavy isotope of oxygen $\delta^{18}$O (b) as the function of the depth and time in ice core obtained at the station "Vostoc" [24].

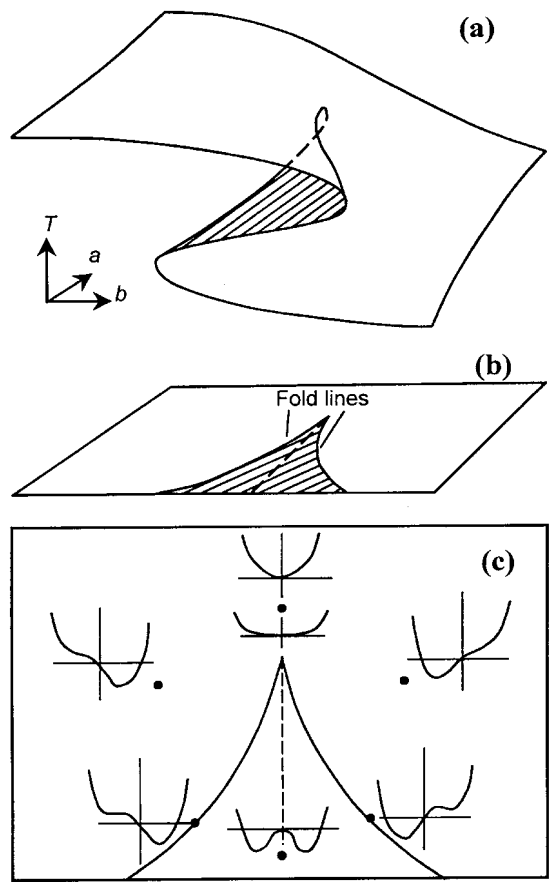

Fig. 2. Canonical form of the variety of fold catastrophe (a), bifurcation set of the critical points (fold line) (b) and form of "potential" at the neighborhood of some points in the plane of control parameters (c).

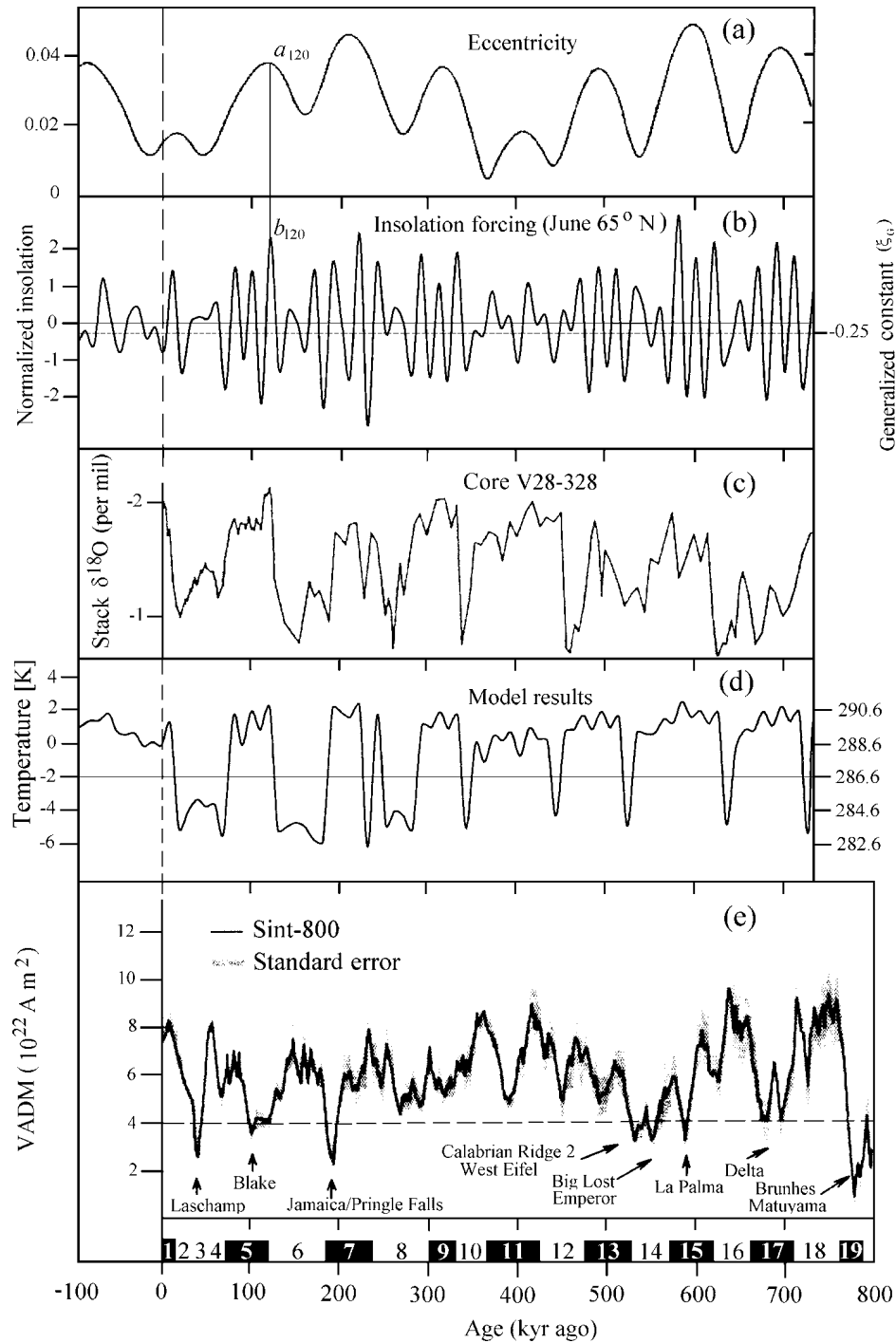

Fig. 3. Model of climatic response to orbital and insolation variations compared with isotopic data on climate of the past 730 kyr and variations of VADM over the past 800 kyr.

Variations in orbital eccentricity (a) [6] and insolation (b) at 65°N at the summer solstice [6] over the past 730 kyr and over the future 100 kyr; (c) oxygen isotope curve V28-328 [34] from the Pacific Ocean (PDB standard) on the time scale of Kominz et. al. [35]; (d) output of our model (18): evolution of the increment of temperature $\Delta T$ relative to the average temperature $T_0$=286.6 K over the past 730 kyr and 100 kyr in future; (e) the synthetic record of VADM variations with standard error, which was obtained by the comparison of 33 records of paleointensity [36]. The horizontal hatched line corresponds to the critical value of intensity, below which the digressions of geomagnetic field were observed (in figure they indicated by the arrow).

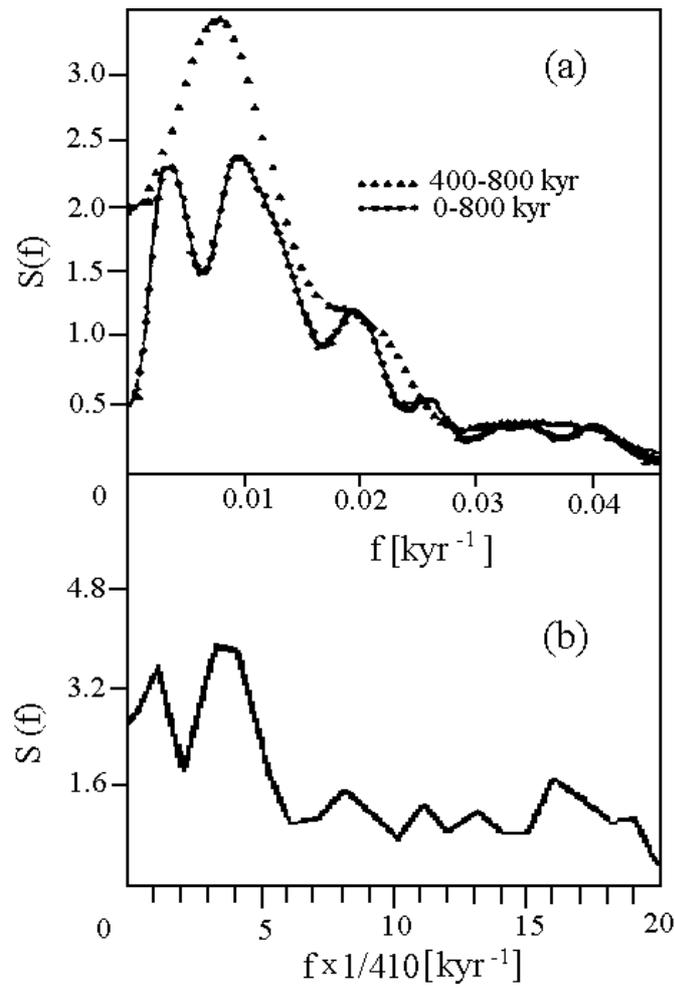

Fig.4. Spectra of the variations of VADM with sliding time window of 800 kyr (a) [36] and variations of the increment of temperature $\Delta T$ with sliding time window of 410 kyr (b). In both spectra, the 100-kyr and 410-kyr periodicity appears around 730 kyr and 800 kyr before present.

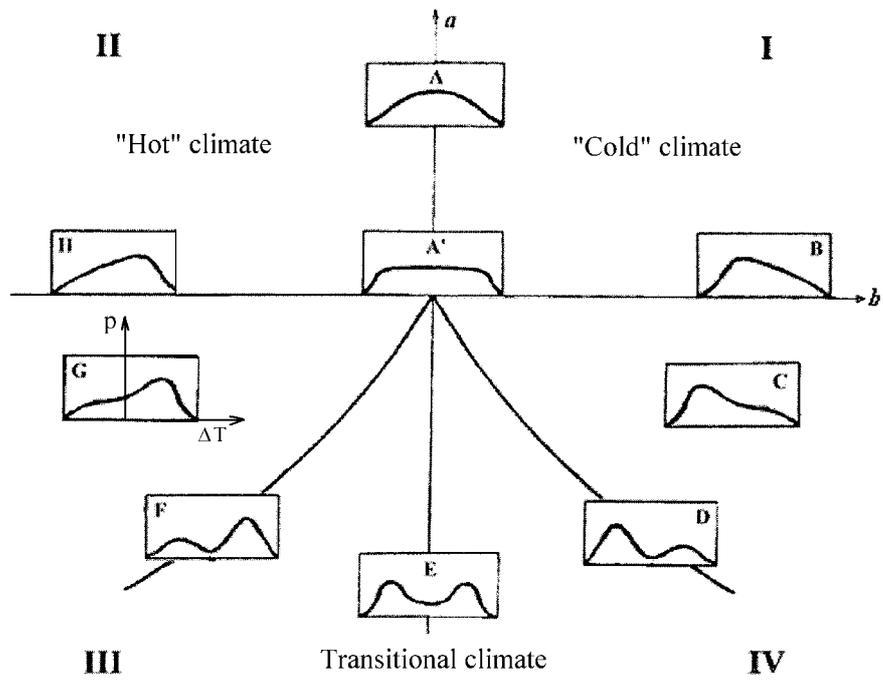

Fig. 5. Changes of probabilistic distribution $p(\Delta T)$ of the increment of temperature $\Delta T$ in the plane of control parameters $a$ a and $b$.

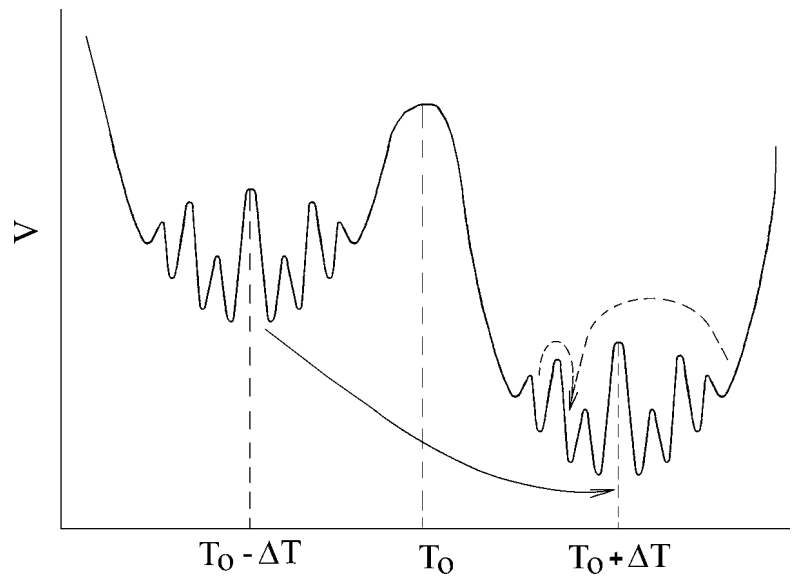

Fig. 6. The hierarchical structure of potential of fold catastrophe. The firm line corresponds to global transition (climate), the hatch lines correspond to local transitions (weather).